\documentclass[letterpaper]{article} 
\usepackage{aaai23}  
\usepackage{times}  
\usepackage{helvet}  
\usepackage{courier}  
\usepackage[hyphens]{url}  
\usepackage{graphicx} 
\usepackage{natbib}  
\usepackage{caption} 
\usepackage{algorithm}
\usepackage{algorithmic}

\usepackage{amsmath,amsfonts}
\usepackage{array}
\usepackage[caption=false,font=normalsize,labelfont=sf,textfont=sf]{subfig}
\usepackage{textcomp}
\usepackage{stfloats}
\usepackage{verbatim}
\usepackage{cite}
\usepackage{microtype}
\usepackage{booktabs}
\usepackage{multirow}
\usepackage{multicol}
\usepackage{makecell}
\usepackage{enumerate}

\usepackage{color}

\usepackage{newfloat}
\usepackage{listings}
\DeclareCaptionStyle{ruled}{labelfont=normalfont,labelsep=colon,strut=off} 
\floatstyle{ruled}
\newfloat{listing}{tb}{lst}{}
\floatname{listing}{Listing}
\title{COLA: Improving Conversational Recommender Systems by \\ Collaborative Augmentation}

\author{
    Dongding Lin\thanks{Equal contribution.},
    Jian Wang\footnotemark[1],
    Wenjie Li
}
\affiliations{
    The Hong Kong Polytechnic University \\
    \{dongding88.lin, jian-dylan.wang\}@connect.polyu.hk,
    cswjli@comp.polyu.edu.hk
}

\begin{document}

\maketitle
\begin{abstract}
Conversational recommender systems (CRS) aim to employ natural language conversations to suggest suitable products to users. Understanding user preferences for prospective items and learning efficient item representations are crucial for CRS. Despite various attempts, earlier studies mostly learned item representations based on individual conversations, ignoring item popularity embodied among all others. Besides, they still need support in efficiently capturing user preferences since the information reflected in a single conversation is limited. Inspired by collaborative filtering, we propose a collaborative augmentation (COLA) method to simultaneously improve both item representation learning and user preference modeling to address these issues. We construct an interactive user-item graph from all conversations, which augments item representations with user-aware information, i.e., item popularity. To improve user preference modeling, we retrieve similar conversations from the training corpus, where the involved items and attributes that reflect the user's potential interests are used to augment the user representation through gate control. Extensive experiments on two benchmark datasets demonstrate the effectiveness of our method. Our code and data are available at \url{https://github.com/DongdingLin/COLA}.
\end{abstract}

\section{Introduction}
With the rapid development of conversational systems, it is expected that recommender systems can employ natural language conversations to provide high-quality recommendations, namely Conversational Recommender Systems (CRS) \cite{christakopoulou2016towards, li2018towards}. Many researchers \cite{liao2019deep, lei2020interactive, gao2021advances, jannach2021survey, ren2021learning, zhou2022c2} have been attracted to explore CRS due to its high impact on e-commerce.

In the area of CRS, tremendous studies \cite{zhang2019text, zhou2020improving, lu2021revcore, deng2021unified, chen2019towards, wang2021finetuning, zhou2022c2} focus on how to learn good item representations and capture user preferences effectively expressed in natural language conversations. The proposed approaches include reinforcement learning \cite{zhang2019text, deng2021unified}, pretraining-finetuning \cite{chen2019towards, wang2021finetuning}, contrastive learning \cite{zhou2022c2}, and so forth. 
However, these studies still suffer from several limitations. \textbf{First}, they mainly learn item representations based on individual conversations, ignoring the attribute of item popularity embodied among many conversations (users). It causes the system not effective enough to distinguish between popular items and others that may be only favored by a small group of users. \textbf{Second}, they still struggle to capture user preferences efficiently since the information in a single conversation that can reflect the user's interests in potential items is limited. To alleviate the above drawbacks, we expect to learn item representations and model user preferences collaboratively by considering their mutual impact. However, it is non-trivial since user-item interaction data is not directly available in CRS.

In this work, we propose a \textbf{COL}laborative \textbf{A}ugmentation (\textbf{COLA}) method to improve CRS, inspired by collaborative filtering \cite{SchaferFHS07}.
Here, the item representations are augmented with user-aware information, i.e., the popularity embodied among all conversations. Similarly, the user representations are augmented with item-aware information that may reveal the user's potential interests. In particular, we first construct an interactive user-item graph by extracting the user-item pairs with liked or disliked relations involved in all conversations from the training corpus. The graph is encoded by an R-GCN \cite{schlichtkrull2018modeling}, where each item node aggregates liked and disliked information (i.e., item popularity) from different user nodes through message passing. We adopt the encoded output to augment item representations with an add operation. In light of the fact that in CRS, similar users converse with the system and provide similar feedback about items and attributes, we use the current conversation to retrieve similar conversations (i.e., similar users) from the training corpus using BM25 \cite{schutze2008introduction}. Then, the contained items and related attributes in the top-$n$ conversations that reveal the user's potential interests are aggregated through a self-attention mechanism. We employ the aggregated output to enhance the user representations with gate control. 

Based on user-item collaboratively-augmented representations, we make recommendations by matching the user with candidate items and selecting the top-$k$ items as the recommended set. In the end, we adopt a widely-used backbone conversation generation model, following existing works \cite{liu2020towards, zhou2022c2}, to generate an appropriate utterance in response to the user.

We summarize the contributions of this paper as follows:
\begin{itemize}
    \item We propose a collaborative augmentation method to help learn item representations and model user preferences, which is simple but effective in improving existing CRS.
    \item We construct an interactive user-item graph to introduce popularity-aware information for item representation learning and propose a retrieval-enhanced approach for user preference modeling.
    \item Extensive experiments on two CRS benchmark datasets demonstrate that our method effectively performs better than various strong baseline models.
\end{itemize}

\section{Related Work}
Conversational Recommender Systems (CRS) \cite{jannach2021survey, sun2018conversational} have been an emerging research topic to provide high-quality recommendations through natural language conversations with users. In order to advance the research, various datasets have been released, such as REDIAL \cite{li2018towards}, TG-REDIAL \cite{zhou2020towards}, INSPIRED \cite{hayati2020inspired}, DuRecDial \cite{liu2020towards, liu2021durecdial}, etc. Existing works are divided into recommendation-biased CRS \cite{christakopoulou2016towards, sun2018conversational, zhou2020leveraging, li2021seamlessly, ren2021learning, xie2021comparison, zhou2020improving, zhou2020towards} and dialogue-biased CRS \cite{li2018towards, chen2019towards, liu2020towards, ma2020cr}. In this paper, we focus on the recommendation-biased CRS. 

Existing recommendation-biased CRSs mainly focus on how to learn item representations and model user preferences. To this end, various approaches have been proposed, including reinforcement learning \cite{zhang2019text, deng2021unified}, pretraining-finetuning \cite{chen2019towards, wang2021finetuning} and contrastive learning \cite{zhou2022c2}. To improve item representation learning, \citet{li2018towards} and \citet{zhou2020improving} introduced domain knowledge graphs and modeled items in the graphs with R-GCN \cite{schlichtkrull2018modeling}. For the user preferences modeling, \citet{XuMLLSY20} created the MGCConvRex corpus to help perform user memory reasoning. \citet{XuYXG0W21} adapted both attribute-level and item-level feedback signals to identify the user's attitude towards an item more precisely. Besides, \citet{li2022user} focused on user-centric conversational recommendations with multi-aspect user modeling, e.g., the user's current conversation session and historical conversation sessions. For the recommendation, there are several studies that focus on asking item attributes \cite{lei2020interactive}, clarifying the user's requests \cite{ren2021learning} and recommendation strategies \cite{zhou2020leveraging, ma2020cr}.

\section{Preliminaries}

\paragraph{Task Definition}
Formally, let $\mathcal{I}={\{{e}_{i}\}}_{i=1}^{m}$ denote the entire item set, $\mathcal{H}={\{{s}_{t}\}}_{t=1}^{n}$ denote a conversation consisting of a list of utterances between a user  (recommendation seeker) and a conversational recommender system, where ${s}_{t}$ is produced by either the user or the system.
Given a conversation history, a CRS aims to select a set of candidate items $\mathcal{I}_t$ that satisfy the user's requests from the entire item set, and then produce an appropriate utterance in response to the user. The system is required to capture user preferences expressed in historical utterances. In some cases,  $\mathcal{I}_t$ may be empty because the system needs to ask questions about item attributes to clarify the user's requests or interests.

\paragraph{Fundamental Framework} Existing methods \cite{chen2019towards, liao2019deep, liu2020towards, lu2021revcore, zhou2022c2} mainly divide CRS into two major modules: \textit{recommendation module} and \textit{response generation module}. The \textit{recommendation module} aims to capture user preferences based on the conversation history and recommend suitable items accordingly. Based on the output of the \textit{recommendation module}, the \textit{response generation module} is utilized to generate natural language responses to interact with users. In general, the recommendation module is critical for the whole performance of CRS.

Many prior works \cite{zhou2020improving, zhang2021kecrs, lu2021revcore, zhou2022c2} have introduced external item-oriented knowledge graphs (KGs) to enrich the learned item representations. For example, one of the widely-used KGs is DBpedia \cite{auer2007dbpedia}, which provides structured knowledge facts about items. Each knowledge fact is formatted as a triple $\langle e_1, r_{k}, e_2\rangle$, where $e_1, e_2 \in \mathcal{E}$ denote entities (or items), $r_{k}$ denotes their relation. To better understand user preferences expressed in natural language conversations, several works \cite{zhou2020improving, lu2021revcore} have introduced word-oriented or commonsense KGs, i.e., ConceptNet \cite{speer2017conceptnet}. The ConceptNet provides commonsense relations between words (e.g., the antonyms relation between ``cheerful'' and ``miserable''), which helps align semantics between word-level information in the conversations and entity-level information in item-oriented knowledge graphs. In ConceptNet, semantic facts are also stored in the form of $\langle w_1, r_c, w_2 \rangle$, where $w_1, w_2 \in \mathcal{V}$ are words, $r_c$ denotes the relation between $w_1$ and $w_2$.

Following the above studies, we also adopt an item-oriented KG and a word-oriented KG as the external data to build our base CRS model. On top of that, we augment the item and user representations with our proposed approach.

\section{Proposed Method}

\begin{figure*}[th!]
    \centering
    \includegraphics[width=0.85\linewidth]{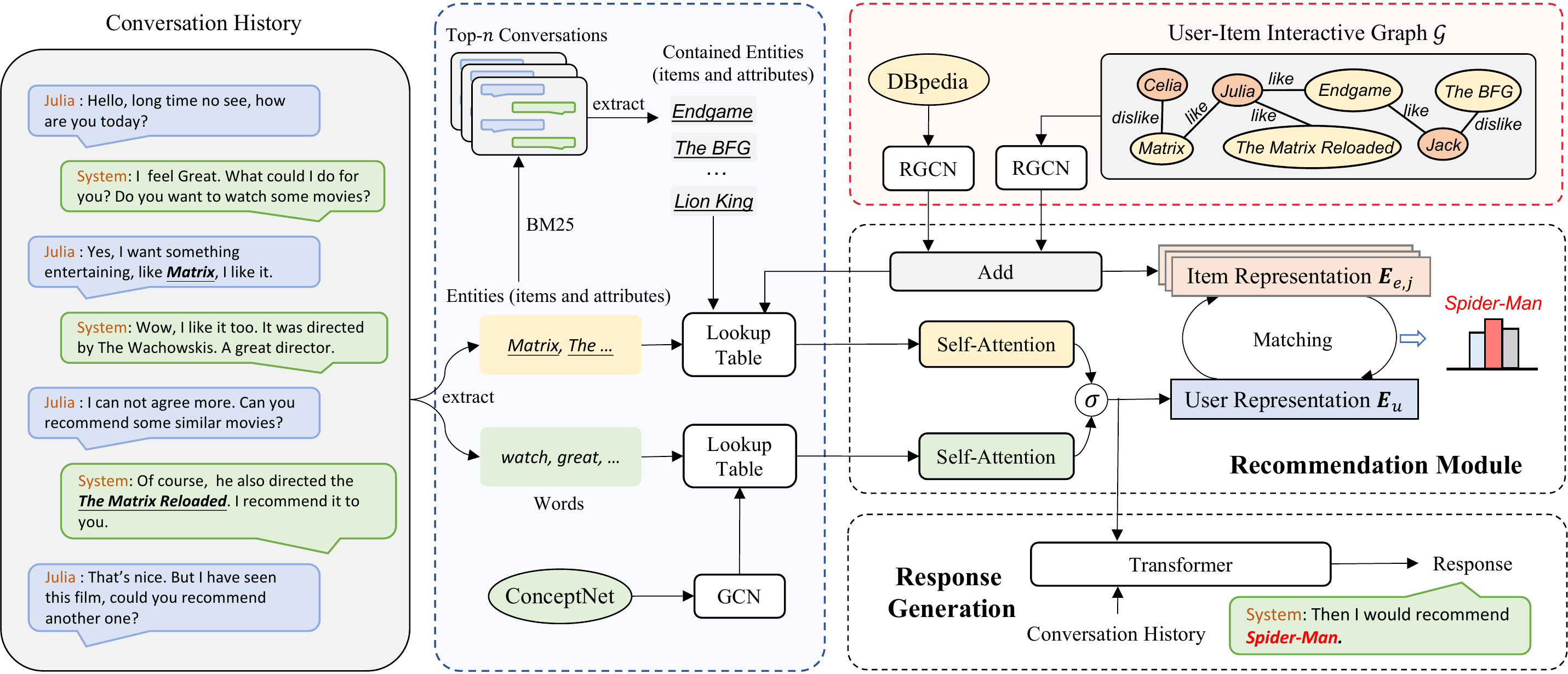}
    \caption{Overview of our proposed collaborative augmentation (COLA) method for CRS.}
    \label{fig:architecture}
    
\end{figure*}

In this section, we propose a \textbf{COL}laborative \textbf{A}ugmentation (\textbf{COLA}) method for CRS, the overview of which is shown in Figure \ref{fig:architecture}.
The key components of our COLA are highlighted in orange and blue dotted boxes, which stand for popularity-aware item representation augmentation and retrieval-enhanced user preference augmentation, respectively. Based on collaboratively augmented user and item representations, we follow the existing fundamental framework for recommendation and response generation.

\subsection{Popularity-aware Item Representation Augmentation}
Before we introduce item representation augmentation, we first adopt an item-oriented KG (e.g., DBpedia) to obtain base item representations following many existing works \cite{chen2019towards, zhou2020improving, lu2021revcore}. Specifically, we employ an R-GCN \cite{schlichtkrull2018modeling} to encode the item-oriented knowledge graph $\mathcal{K}$. Formally, the representation of each entity $e$ in the $\mathcal{K}$ is calculated as follows:
\begin{align}
    \mathbf{k}_{ e}^{\ell+1}=\text{ReLU}(\sum_{r\in\mathcal{R}_{k}}\sum_{e' \in {\mathcal{E}^{r}_{e}}}\frac{1}{Z_{e, r}}\mathbf{W}_{k,r}^{\ell}{\mathbf{k}}^{\ell}_{e'}+\mathbf{W}_{k}^{\ell}\mathbf{k}_{e}^\ell),
    \label{eq1}
\end{align}
where $\mathbf{k}_{e}^{\ell} \in \mathbb{R}^d$ is the node representation of $e$ at the $\ell$-th layer, $d$ is the embedding dimension. $\mathcal{E}^{r}_{e}$ denotes the set of neighboring nodes for $e$ under the relation $r$. $\mathbf{W}_{k,r}^{\ell}$ is a learnable relation-specific transformation matrix for the embeddings from neighboring nodes with relation $r$, $\mathbf{W}_{k}^{\ell}$ is a learnable matrix for transforming the representations of nodes at the $\ell$-th layer. $Z_{e, r}$ is a normalization factor.  ReLU($\cdot$) is an activation function. After aggregating the KG information, we adopt the output of R-GCN in Eq. \ref{eq1} as the base item representations, denoted as $\mathbf{E}_{\mathcal{K}}=\{\mathbf{k}_{e, 1}, \mathbf{k}_{e, 2}, \cdots, \mathbf{k}_{e, m}\}$, where $m$ denotes the number of items.

More importantly, we aim to augment the base item representations with user-aware information, i.e., the popularity embodied among all conversations. Since natural language conversations between a user and the system contain specific key words (e.g., ``like'', ``enjoy'', ``dislike'', ``hate'', etc.) that reflect the user's attitudes towards items and related attributes, we extract all the user's liked and disliked items mentioned in user-system conversations. Based on the training corpus, we obtain various user-item pairs with liked or disliked relation and then construct an interactive user-item graph $\mathcal{G}$ (see Figure \ref{fig:architecture}) accordingly. In the graph $\mathcal{G}$, each edge between a user node and an item node represents the user's preferences (i.e., like or dislike), given by:
\begin{align}
\label{IG}
\mathcal{G}=\langle u, r, e \rangle, u \in \mathcal{U}, e \in \mathcal{I}, r \in \mathcal{R}_{\mathcal{G}},
\end{align}
where $\mathcal{U}$ denotes all users, $\mathcal{I}$ denotes the entire item set, $\mathcal{R}_{\mathcal{G}}=\{\text{``like''}, \text{``dislike''}\}$. Intuitively, the item nodes liked by more users (or disliked by fewer users) are more popular than those liked by fewer users (or disliked by more users). 

To obtain popularity-aware item representations, we adopt another R-GCN to encode the user-item graph $\mathcal{G}$ as follows:
\begin{align}
    \label{eq3}
    \mathbf{v}_{\mathcal{G},e}^{\ell+1}=\text{ReLU}(\sum_{r\in\mathcal{R}_{\mathcal{G}}}\sum_{u \in {\mathcal{G}^{r}_{e}}}\frac{1}{Z_{e, r}}\mathbf{W}_{\mathcal{G},r}^{\ell}{\mathbf{v}}^{\ell}_{u}+\mathbf{W}_{\mathcal{G}}^{\ell}\mathbf{v}_{\mathcal{G},e}^\ell), \\
    \label{eq4}
    \mathbf{v}_{\mathcal{G},u}^{\ell+1}=\text{ReLU}(\sum_{r\in\mathcal{R}_{\mathcal{G}}}\sum_{e \in {\mathcal{G}^{r}_{u}}}\frac{1}{Z_{u, r}}\mathbf{W}_{\mathcal{G},r}^{\ell}{\mathbf{v}}^{\ell}_{e}+\mathbf{W}_{\mathcal{G}}^{\ell}\mathbf{v}_{\mathcal{G},u}^\ell),
\end{align}
where $\mathbf{v}_{\mathcal{G},e}, \mathbf{v}_{\mathcal{G},u} \in \mathbb{R}^{d}$ is the $\ell$-th layer's representation of item $e$ and user $u$, respectively. $\mathcal{G}^r_e, \mathcal{G}^r_u$ are the one-hop neighbor set of $e$ and $u$ under the relation $r$. $\mathbf{W}_{\mathcal{G}, r}^\ell$ and $\mathbf{W}_{\mathcal{G}}^\ell$ are trainable weights of layer $\ell$. $Z_{e, r}$ and $Z_{u, r}$ are normalization factors. According to Eq. \ref{eq3} and Eq. \ref{eq4}, each item node aggregates liked and disliked information (i.e., item popularity) from different user nodes through message passing. After computation, we adopt the item representations based on Eq. \ref{eq3} as the popularity-aware item representations, denoted as $\mathbf{E}_{\mathcal{G}}=\{\mathbf{v}_{e, 1}, \mathbf{v}_{e, 2}, \cdots, \mathbf{v}_{e, m}\}$, where $m$ denotes the number of items. Finally, we augment the base item representations $\mathbf{E}_{\mathcal{K}}$ with popularity-aware item representations $\mathbf{E}_{\mathcal{G}}$ by an element-wise add operation:
\begin{align}
    \mathbf{E}_{e, j} = \mathbf{k}_{e, j} + \mathbf{v}_{e, j}
\end{align}
where $\mathbf{k}_{e, j} \in \mathbf{E}_{\mathcal{K}}$, $\mathbf{v}_{e, j} \in \mathbf{E}_{\mathcal{G}}$, $j \in [1, m]$.

\subsection{Retrieval-enhanced User Preference Augmentation}
Prior to introducing user preference augmentation, we first briefly describe the basics of user preference modeling. The user's feedback about specific items is essential to capturing user preferences from natural language conversations.
To this end, we extract the mentioned items involved in the conversation history and collect the representations of these items from the entire augmented item representations $\mathbf{E}_e$ through a lookup operation, obtaining $\mathbf{E}_{(m)} \in \mathbb{R}^{{\ell^m} \times d}$, where ${\ell^m}$ denotes the number of mentioned items in the conversation history, $d$ denotes the embedding dimension. Then, to better understand user preferences expressed in natural language conversations, we adopt a word-oriented KG, i.e., ConceptNet, to help align semantics between word-level information in conversations and entity-level information in the item-oriented KG following existing works \cite{zhou2020improving, lu2021revcore}. In detail, we utilize a GCN  \cite{zhang2019graph} to encode the ConceptNet $\mathcal{C}$. The representation of each word $w$ in the $\mathcal{C}$ is calculated as follows:
\begin{align}
    \mathbf{V}_c^{\ell+1}=\text{ReLU}(\mathbf{D}^{-\frac{1}{2}}\mathbf{A}\mathbf{D}^{-\frac{1}{2}}\mathbf{V}_c^{\ell}\mathbf{W}_c^{\ell+1}),
\end{align}
where $\mathbf{V}_c^\ell \in \mathbb{R}^{\mathbf{V}\times d}$ are the representations of nodes and $\mathbf{W}_c^{\ell+1}$ is a learnable matrix at the $\ell+1$-th layer. $\mathbf{A}$ is the adjacency matrix of the graph and $\mathbf{D}$ is a diagonal degree matrix with entries $\mathbf{D}\lbrack i, i \rbrack=\Sigma_j\mathbf{A} \lbrack i, j \rbrack$. We then adopt a simple lookup operation to obtain the representations of the contextual words (except stop words) in the conversation history that also appeared in the ConcepNet $\mathcal{C}$, denoted as $\mathbf{C}_{(m)} \in \mathbb{R}^{{\ell^c} \times d}$, where ${\ell^c}$ denotes the length of the conversation history, $d$ denotes the embedding dimension. 

With the obtained representations  $\mathbf{E}_{(m)}$ and  $\mathbf{C}_{(m)}$, we investigate how to augment them with item-aware information that may reveal the user's potential interests. Intuitively, similar users converse with the system and provide similar feedback about items and attributes. As shown in Figure \ref{fig:architecture}, we use the current conversation to retrieve similar conversations (i.e., similar users) from the training corpus using BM25 \cite{schutze2008introduction}. Specifically, let all conversations in the training corpus be formed as $D=\{\mathcal{H}_{i}\}^{N}_{i=1}$, where $N$ denotes the total number of conversations. We extract the entities (items and attributes) involved in the current conversation as the query $Q$, then use the BM25 algorithm to retrieve similar conversations from $D$. We select the top-$n$ conversations according to the similarity score between $Q$ and each conversation in $D$. The contained entities (items and attributes) in the top-$n$ conversations are denoted as $E_{(r)}$. The above process is formulated as follows:
\begin{align}
E_{(r)} = \text{Top-}n(\text{BM25}(Q, \{\mathcal{H}_{i}\}^{N}_{i=1}),
\end{align}
Similarly, we collect the representations of  $E_{(r)}$ from $\mathbf{E}_{e}$ through a lookup operation, denoted as $\mathbf{E}_{(r)}$.

Given the representations $\mathbf{C}_{(m)}$, $\mathbf{E}_{(m)}$, and $\mathbf{E}_{(r)}$, we introduce how to incorporate them together to obtain the augmented user preference representations. First, since $\mathbf{E}_{(m)}$ and $\mathbf{E}_{(r)}$ are calculated through the same semantic space, we combine them with a concatenation operation and then apply a self-attention operation to capture the item-oriented user preferences, which is given by:
\begin{align}
    \mathbf{E}_{(mr)} &= [\mathbf{E}_{(m)}; \mathbf{E}_{(r)}], \\
    \alpha &= \text{softmax}(\mathbf{b}_{1}^\top \cdot \text{tanh}(\mathbf{W}_{1}\mathbf{E}_{(mr)})), \\
    \mathbf{v}_{\Tilde{e}} &= \alpha\cdot\mathbf{E}_{(mr)} 
\end{align}
where $[;]$ denotes the concatenation operation, $\mathbf{W}_{1}$ is a learnable parameter matrix, and $\mathbf{b}_{1}$ is a learnable bias. Second, we employ a similar self-attention operation on $\mathbf{C}_{(m)}$ to better capture the semantics of the contextual words in the conversations, which is given by:
\begin{align}
    \beta &= \text{softmax}(\mathbf{b}_{2}^\top \cdot \text{tanh}(\mathbf{W}_{2}\mathbf{C}_{(m)})), \\
    \mathbf{v}_{\Tilde{w}} &= \beta\cdot\mathbf{C}_{(m)}
\end{align}
where $\mathcal{\beta}$ denotes the attention weights reflecting the importance of each word, $\mathbf{W}_{2}$ is a learnable parameter matrix, and $\mathbf{b}_{2}$ is a learnable bias. 

In the end, we fuse the item-oriented user representation and word-oriented user representation through a gate, obtaining the ultimate user preference representation $\mathbf{E}_{u}$, which is given by:
\begin{align}
    \mathbf{E}_{u} &= \gamma \cdot \mathbf{v}_{\Tilde{e}} + (1-\gamma) \cdot \mathbf{v}_{\Tilde{w}}, \\
    \gamma &= \sigma(\mathbf{W}_{3}[\mathbf{v}_{\Tilde{e}}; \mathbf{v}_{\Tilde{w}}]),
\end{align}
where $\mathbf{W}_{3}$ is a learnable parameter matrix, $\sigma$ denotes the Sigmoid activation function. $[;]$ denotes the concatenation operation.

\subsection{Recommendation}
Given the augmented user representation $\mathbf{E}_{u}$ and all augmented item representations $\mathbf{E}_e$, we compute the probability that recommends $j$-th item to a user $u$ as follows:
\begin{equation}
    P_{rec}(j) = \text{softmax}(\mathbf{E}_{u}^\top \cdot \mathbf{E}_{e, j}),
    \label{rec_score}
\end{equation}
where $\mathbf{E}_{e, j}$ is the item representation of $j$-th item based on $\mathbf{E}_{e}$. Following existing works \cite{zhou2020improving, zhou2022c2}, we apply cross-entropy loss as the optimization objective to learn the model parameters:
\begin{align}
    \mathcal{L}_{rec}=-\sum_{i=1}^{N}\sum_{j=1}^{M} y_{ij}\log (P_{rec}^{(i)}(j)),
\end{align}
where $N$ is the number of conversations, and $i$ is the index of a conversation. $M$ is the number of items, and $j$ is the index of an item. $y_{ij}$ denotes the item label. During inference, we utilize Eq. \ref{rec_score} to rank all candidate items from the item set $\mathcal{I}$ and select top-$k$  items as the recommended set.

\subsection{Response Generation}
Following existing works \cite{zhou2020improving, lu2021revcore}, we adopt the widely-used language generation model Transformer \cite{vaswani2017attention} as the backbone model for response generation. We first use a Transformer encoder to encode conversation history $\mathcal{H}$, obtaining the hidden representations $\mathbf{H}=(\mathbf{h}_{1},\mathbf{h}_{2},\cdots, \mathbf{h}_{n})$. Then, we adopt a Transformer decoder with the encoder-decoder attention mechanism, where the conditional generation distribution is approximated following \citet{zhou2020improving}:
\begin{align}
   P_{\theta}(y_t|y_{<t})&=\text{softmax}(\mathbf{W}\mathbf{s}_{t}+\mathbf{b}) \\
   \mathbf{s}_{t}&=\text{Transformer}(\mathbf{s}_{t-1}, \mathbf{H}, \mathbf{C}_{(m)}, \mathbf{E}_{(mr)})
\end{align}
where $\mathbf{s}_{t}$ denotes decoder hidden state at $t$-th time step, $\mathbf{W}$ and $\mathbf{b}$ are trainable parameters. We train the Transfomer language model by minimizing the negative log-likelihood as follows:
\begin{align}
    \mathcal{L}_{gen}=-\frac{1}{T}\sum_{t=1}^{T}\log P_{\theta}(y_t|y_{<t}),
\end{align}
where $T$ is the length of the response. During inference, we employ the trained model to generate an appropriate utterance word by word in response to the user.

\section{Experimental Setup}

\subsection{Datasets}
We conduct experiments on two widely-used CRS datasets, namely REDIAL \cite{li2018towards} and TG-ReDial \cite{zhou2020towards}. The REDIAL dataset is collected by crowd-sourced workers and contains
10,006 conversations about English movies. The TG-ReDial dataset is constructed from 10,000 conversations between a user and a recommender in a topic-guided way, with all topics concerning Chinese movies. Both two datasets are split into training/validation/testing sets with a ratio of 8:1:1. Overall, the statistics of the REDIAL and TG-ReDial datasets are shown in Table \ref{statistic}.

\begin{table}[th]
\centering
\resizebox{0.76\linewidth}{!}{
\begin{tabular}{l|r|r}
\toprule
    Dataset       & REDIAL  & TG-ReDial \\
\midrule
\# Users          & 956     & 1,482     \\
\# Conversations & 10,006  & 10,000    \\
\# Utterances    & 182,150 & 129,392   \\
\# Items         & 64,362  & 33,834   \\
\# Words / Utterance & 14.5 & 19.0 \\
\# Items / Conversation & 4.2 & 3.0 \\
\# Turns / Conversation & 15.0 & 12.0 \\
\bottomrule
\end{tabular}}
\caption{Statistics of the REDIAL and TG-ReDial datasets.}
\label{statistic}
\end{table}

For the REDIAL dataset, we directly utilize the annotated relations (``like'' or ``dislike'') regarding the user's attitudes toward an item in each conversation. Then we extract various user-item pairs with like or dislike relations to construct an interactive user-item graph accordingly. 
For the TG-ReDial dataset, we automatically detect the like or dislike relations with a predefined keyword set (e.g., ``love'', ``fear'', etc.) from all conversation utterances. In addition, the items that are accepted by the user at the end of a conversation are viewed as the  ``liked'' items, while the items that are rejected are viewed as ``disliked'' ones. Similarly,  we  construct another interactive user-item graph.

\subsection{Baseline Methods}
We compare our method with several competitive baseline models: (1) \textbf{ReDial} \cite{li2018towards}: It is the benchmark model released in the REDIAL dataset. Basically, it consists of a recommendation module based on auto-encoder \cite{he2017distributed} and a response generation module using HRED \cite{serban2017hierarchical}. (2) \textbf{KBRD} \cite{chen2019towards}: It adopts an item-oriented knowledge graph to improve the semantics of contextual items in conversations and then employs a Transformer \cite{vaswani2017attention} for response generation. (3) \textbf{TG-ReDial} \cite{zhou2020towards}:  It is the benchmark model released in the TG-ReDial dataset, which first predicts topics or items and then generates corresponding responses based on the predicted topics or items. (4) \textbf{KECRS} \cite{zhang2021kecrs}: It introduces an external high-quality KG and is trained with Bag-of-Entity loss to better capture user preferences for conversational recommendation. (5) \textbf{KGSF} \cite{zhou2020improving}: It employs mutual information maximization to align the semantics between words in the conversation and items in the knowledge graph. (6) \textbf{NTRD} \cite{NTRD2021EMNLP}: It decouples dialogue generation from item recommendation via a two-stage strategy, which makes the system more flexible and controllable. (7) \textbf{CR-Walker} \cite{ma2020cr}: It performs tree-structured reasoning on knowledge graphs and generates informative dialog acts to guide response generation. (8) \textbf{RevCore} \cite{lu2021revcore}: It proposes a review-enhanced framework that uses item reviews to improve recommendation and response generation, where the item reviews are selected by sentiment-aware retrieval.

For a fair comparison, we adopt the implementations of all the above models implemented by the open-source  CRS toolkit CRSLab \cite{zhou2021crslab}. 

\begin{table*}[th!]
\centering
\resizebox{0.95\linewidth}{!}{
\begin{tabular}{l|c|c|c|c|c|c|c|c|c|c|c|c}
\toprule
 & \multicolumn{6}{c|}{REDIAL} & \multicolumn{6}{c}{TG-ReDial}  \\
\midrule
Model & R@1 & R@10 & R@50 & MRR@1 & MRR@10 & MRR@50 & R@1 & R@10 & R@50 & MRR@1 & MRR@10 & MRR@50\\
\midrule
ReDial & 0.020 & 0.140 & 0.320 & 0.020 & 0.072 & 0.075 & 0.000 & 0.002 & 0.013 & 0.000 & 0.002 & 0.004 \\
KECRS & 0.021 & 0.143 & 0.340 & 0.021 & 0.075 & 0.080 & 0.002 & 0.026 & 0.069  & 0.002 & 0.003 & 0.004\\
KBRD & 0.031 & 0.150 & 0.336 & 0.031 & 0.074 & 0.078 & 0.005 & 0.032 & 0.077 & 0.005 & 0.004 & 0.006 \\
TG-ReDial & 0.032 & 0.169 & 0.380 & 0.032 & 0.077 & 0.082 & 0.003 & 0.017 & 0.051 & 0.003 & 0.006 & 0.008 \\
NTRD & 0.037 & 0.197 & 0.381 & 0.037 & 0.081 & 0.085 & 0.004 & 0.025 & 0.072 & 0.004 & 0.005 & 0.007 \\
CR-Walker & 0.040 & 0.187 & 0.376 & 0.040 & 0.079 & 0.082 & - & - & - & - & - & - \\

KGSF & 0.039 & 0.183 & 0.378 & 0.039 & 0.082 & 0.091 & \bfseries{0.005} & 0.030 & 0.074  & \bfseries{0.005} & 0.007 & 0.008 \\
RevCore & 0.046 & 0.211 & 0.396 & 0.045 & 0.083 & 0.092 & 0.004 & 0.029 & 0.075 & 0.004 & 0.007 & 0.008\\
\midrule
COLA (Ours) & \bfseries{0.048} & \bfseries{0.221} & \bfseries{0.426} & \bfseries{0.048} & \bfseries{0.086} & \bfseries{0.096}
& \bfseries{0.005} & \bfseries{0.034} & \bfseries{0.079}  & \bfseries{0.005} & \bfseries{0.009} & \bfseries{0.010} \\
\bottomrule
\end{tabular}}
\caption{Evaluation results of recommendation on the REDIAL dataset and TG-ReDial dataset. The best results are highlighted in bold and the improvements are statistically significant compared to baselines ($t$-test with $p$-value $<$ 0.05).}
\label{rec_eval}
\end{table*}

\subsection{Evaluation Metrics}
Following many existing works, we evaluate the performance of a CRS model in terms of both recommendation and response generation. The automatic metrics for recommendation evaluation are Recall@$k$ (R@$k$, $k$= 1, 10, 50) and Mean Reciprocal Rank@$k$ (MRR@$k$, $k$= 1, 10, 50), which evaluate whether the model's recommended top-$k$ items hit the ground truth items provided by human recommenders. 
The evaluation of response generation includes automatic and human evaluations. For automatic evaluation, we adopt the perplexity (PPL), BLEU-2,3 \cite{papineni2002bleu} and Distinct $n$-gram (DIST-$n$, $n$ = 2, 3, 4) \cite{chen2019towards,zhou2020improving}. The perplexity is a measurement for the fluency of natural
language, where lower perplexity refers to higher fluency. The BLEU-2,3 measure word overlaps of the generated responses and the ground truth responses. The DIST-$n$ measures the diversity of the generated responses at the sentence-level.
For human evaluation, we recruit three annotators to evaluate the generated responses manually. The annotators are required to rate a score in the range \{0, 1, 2\} to each generated response from \textit{fluency} and \textit{informativeness}, following \cite{zhou2020improving,zhang2021kecrs}. We calculate the Fleiss's kappa \cite{fleiss1971measuring} to measure the inter-annotator agreement and adopt the average score of three annotators as the human evaluation result.

\subsection{Implementation Details}
We implement our approach with Pytorch. The hidden dimension is set to 128 and 300 for the recommendation module and response generation module, respectively. For the BM25 algorithm, the number of the top conversations $n$ is set to 1. The max length of conversation history is limited to 256. For both R-GCN and GCN, the number of layers is set to 2 in consideration of efficacy and efficiency. The embeddings of both RCGN and GCN are randomly initialized. The normalization factor of R-GCN is set to 1.0 by default. We use the pretrained 300-$d$ word2vec \cite{MikolovSCCD13} embeddings\footnote{\url{https://radimrehurek.com/gensim/models/word2vec.html}} for Transformer during response generation. During training, we use the Adam \cite{kingma2014adam} optimizer with an initial learning rate of 0.001 and a gradient clipping strategy to restrict the gradients within [0, 0.1]. The batch size is set to 256.  We train our model with 30 epochs for both recommendation and response generation. During testing, we use the greedy search algorithm to generate responses, where the max decoding length is set to 30.

\section{Results and Analysis}

\subsection{Evaluation on Recommendation}
The evaluation results of recommendation on the two datasets are reported in Table \ref{rec_eval}. The best result in terms of the corresponding metric is highlighted in boldface. The CR-Walker is not evaluated on the TG-ReDial dataset owing to a lack of pre-constructed reasoning trees.
As shown in Table \ref{rec_eval}, the ReDial model performs inferior compared with other models since it utilized no external KG to enrich the item representations. By using an item-oriented KG (i.e., DBpedia) as external knowledge, KECRS achieves more accurate recommendations. Both KBRD and KGSF achieve significant improvements since they integrate recommendation and response generation through joint learning. Besides, RevCore performs better than other baseline models by utilizing item reviews to enhance item representations.

As shown in Table \ref{rec_eval}, we observe that our COLA outperforms all baseline models in terms of all metrics. For example, our model obtains 4.5\% and 7.5\% improvements compared to RevCore in terms of R@10 and R@50 on the REDIAL dataset, which shows our augmented item representations and user representations benefit the model to make more accurate recommendations. For the rank of the recommended items, our model also outperforms other models as shown by the MRR metrics on both two datasets. Additionally, we observe that all models obtain much lower recall scores on the TG-ReDial dataset compared to that on the REDIAL dataset. The reason is that the contextual items in conversations from the TG-ReDial dataset are much sparser than those in the REDIAL dataset, making it more challenging to capture user preferences and item representations and make recommendations accurately.

\subsection{Ablation Study}
We conducted an ablation study based on different variants of our model on the two datasets to verify the effectiveness of each component. We focus on the following components and set them for ablation experiments accordingly: (1) without the interactive user-item graph (w/o IG); (2) without the retrieved top-$n$ conversations (w/o RT); (3) without the item-oriented KG (w/o DB), i.e., DBpedia; (4) without the word-oriented KG (w/o CN), i.e., ConceptNet.

From the ablation study results reported in Table \ref{ablation}, we observe that each component contributes to making more accurate recommendations. In particular, the performance of COLA w/o DB drops sharply in both two datasets in terms of all metrics. It shows that the external item-oriented KG is essential since it contains items' rich attributes that benefit the system to represent the items more effectively. Besides, the R@10 of COLA w/o IG declines by 4.5\% on the ReDial dataset, demonstrating that the item popularity-aware information benefits the system to augment item representations and make more accurate recommendations. The R@10 of COLA w/o RT on the REDIAL dataset deceases by 2.7\%, which indicates that the retrieved conversations help the model to capture user preferences better and recommend appropriate items accordingly. The results of COLA w/o CN on two datasets verify that introducing a commonsense graph helps recommendation since it aligns the semantics between word-level information in conversations and entity-level information in item-oriented knowledge graphs.

\begin{table}[t!]
\centering
\resizebox{0.98\linewidth}{!}{
\begin{tabular}{c|l|cccc}
\toprule
 & Model & R@10 & R@50 & MRR@10 & MRR@50 \\
\midrule
\multirow{5}{*}{REDIAL} & COLA & \bfseries{0.221} & \bfseries{0.426} &  \bfseries{0.086} & \bfseries{0.096} \\
& \ \ w/o IG & 0.211 & 0.415 & 0.083 & 0.091 \\ & \ \  w/o RT & 0.215 & 0.417 & 0.084 & 0.092 \\ & \ \  w/o DB &  0.145 & 0.327 & 0.073 & 0.076 \\ & \ \  w/o CN & 0.218 & 0.420 & 0.084 & 0.094\\ 
\midrule
\multirow{5}{*}{TG-ReDial} & COLA & \bfseries{0.021} & \bfseries{0.060} &  \bfseries{0.009} & \bfseries{0.010} \\
& \ \ w/o IG & 0.019 & 0.058 & 0.007 & 0.008 \\ & \ \  w/o RT & 0.020 & 0.058 & 0.008 & 0.008 \\ & \ \  w/o DB & 0.014 & 0.049 & 0.004 & 0.005 \\ & \ \  w/o CN & 0.020 & 0.059 & 0.007 & 0.008 \\ 
\bottomrule
\end{tabular}}
\caption{Ablation results of recommendation on the REDIAL dataset and TG-ReDial dataset. 
}
\label{ablation}
\end{table}

\subsection{Evaluation on Response Generation}

\paragraph{Automatic Evaluation}

\begin{table*}[th!]
\centering
\resizebox{0.88\linewidth}{!}{
\begin{tabular}{l|c|c|c|c|c|c|c|c|c|c|c|c}
\toprule
 & \multicolumn{6}{c|}{REDIAL} & \multicolumn{6}{c}{TG-ReDial}  \\
\midrule
Model & PPL &  BLEU-2 & BLEU-3 & DIST-2 & DIST-3 & DIST-4 & PPL & BLEU-2 & BLEU-3 & DIST-2 & DIST-3 & DIST-4 \\
\midrule
ReDial & 28.1 & 0.021 & 0.007 & 0.225 & 0.236 & 0.228  & 81.6 & 0.020 & 0.005 & 0.070 & 0.121 & 0.137 \\
KECRS & 19.3 & 0.014 & 0.005 & 0.232 & 0.310 & 0.382  & 31.7 &  0.015 & 0.003 & 0.072 & 0.133 & 0.145 \\
KBRD & 17.9 & 0.022 & 0.008 & 0.243 & 0.349 & 0.399 & 28.0 &  0.021 & 0.006 & 0.108 & 0.172 & 0.193 \\
NTRD  & 12.1 &  0.024 & 0.008 & 0.256 & 0.371 & 0.426  & 22.1 &  0.019 & 0.006 & 0.122 & 0.186 & 0.256\\
KGSF & 11.7 & 0.024 & 0.009 & 0.289 & 0.434 & 0.519  & 17.2 &  0.022 & 0.007 & 0.137 & 0.193 & 0.278 \\
RevCore & 10.2 & 0.025 & 0.009 & 0.373 & 0.507 & 0.598  & 15.4 &  0.023 & 0.007 & 0.146 & 0.224 & 0.299 \\
\midrule
COLA (Ours) & \bfseries{8.6} & \bfseries{0.026} & \bfseries{0.012} & \bfseries{0.387} & \bfseries{0.528} & \bfseries{0.625}  & \bfseries{12.8} &  \bfseries{0.025} & \bfseries{0.008} & \bfseries{0.151} & \bfseries{0.238} & \bfseries{0.313} \\
\bottomrule
\end{tabular}}
\caption{Automatic evaluation results of response generation on the REDIAL dataset and TG-ReDial dataset. The best results are highlighted in bold and the improvements are statistically significant compared to baselines ($t$-test with $p$-value $<$ 0.05).}
\label{conv_eval}
\end{table*}

Our automatic evaluation results of response generation on the REDIAL dataset and TG-ReDial dataset are reported in Table \ref{conv_eval}. The best result in terms of the corresponding metric is highlighted in boldface. The results of the two models, TG-ReDial and CR-Walker, are not included in Table \ref{conv_eval} since they utilize a pre-trained language model for generation, which will bring an unfair comparison. On both two datasets, KECRS performs badly in comparison to other models in terms of DIST-$n$ since it tends to produce utterances with repeated words or entities. Besides, in terms of BLUE and DIST metrics, the superior performances of KBRD and KGSF compared to the ReDial model show that external item-oriented KG and semantic similarity information both contribute to generating better responses. Compared to the baseline methods, our COLA achieves improvements over many metrics on the two datasets. It verifies that gains from item recommendations make it more likely to generate appropriate responses with the correct entities.

\paragraph{Human Evaluation}

\begin{table}[t!]
\centering
\resizebox{0.76\linewidth}{!}{
\begin{tabular}{l|cc|cc}
\toprule
Model & Fluency & $\kappa$ & Inform. & $\kappa$ \\ 
\midrule
ReDial & 1.31 & 0.41 & 0.89  & 0.48 \\
KBRD &  1.35 & 0.45 &  1.01  & 0.56 \\
KECRS & 1.40 & 0.47 & 1.27  & 0.51\\
KGSF &  1.81 & 0.52 & 1.58  & 0.48\\
RevCore  & 1.88  & 0.48 & 1.64 & 0.59 \\
\midrule
COLA (Ours) & \bfseries{1.91} & 0.52 & \bfseries{1.70} & 0.49 \\
\bottomrule
\end{tabular}}
\caption{Human evaluation results on the REDIAL dataset. ``Inform.'' denotes ``informativeness'', $\kappa$ denotes kappa.}
\label{human_eval}
\end{table}

The human evaluation results are reported in Table \ref{human_eval}. The Fless's kappa scores are mainly distributed in [0.4, 0.6], which denotes moderate inter-annotator agreement. We observe that the human evaluation scores of KGSF and RevCore are better than KECRS and KBRD, which demonstrate the effectiveness of using external KGs to bridge the semantic gap between items and natural language utterances. Our COLA performs the best in both metrics compared to all baseline models. By leveraging the augmented item representations and user representations, our model is able to capture user preferences effectively, which assists the model to recommend suitable items and then steer the model to generate more informative words and maintain the fluency of the generated responses.

\subsection{Case Study}

\begin{figure}[th!]
\includegraphics[width=0.95\linewidth]{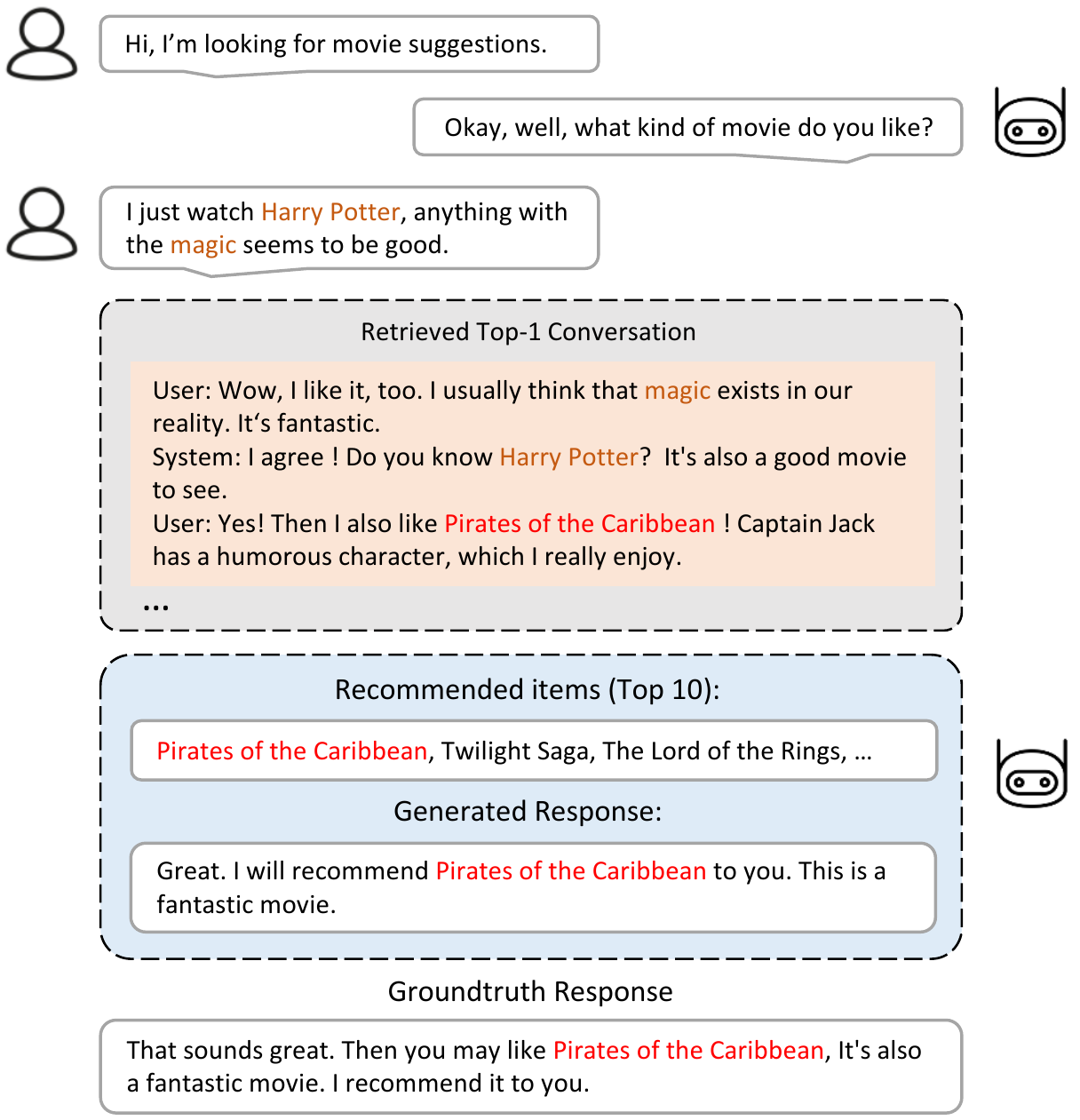}
\centering
\caption{An illustrative case produced by our COLA on the REDIAL dataset.}
\label{fig:case}
\end{figure}

To present our COLA's recommendation and response generation qualities, we show an illustrative case in Figure \ref{fig:case}. We observe that in the beginning, the user asks for a movie recommendation, and the system generates a question to clarify the user's interests. After obtaining the feedback that indicates the user likes a ``\textit{magic}'' movie and is a fan of ``\textit{Harry Potter}'', our COLA calculates the similarity between conversation history and the corpus with our retrieval-enhanced user preference augmentation module. With the top-$1$ retrieved conversation, our model better captures the user's potential interests, including the movie ``\textit{Pirates of the Caribbean}''. Therefore, the system is able to recommend ``\textit{Pirates of the Caribbean}'' with a higher probability. Ultimately, our model generates an appropriate response to complete the recommendation accordingly.

\section{Conclusion}
In this paper, we propose a collaborative augmentation (COLA) method to improve conversational recommender systems. Our COLA aims to augment item representations with user-aware information and user preference representations with item-aware information. In particular, we construct an interactive user-item graph to introduce popularity-aware information for item representation learning and propose a retrieval-enhanced approach for user preference modeling. Experimental results demonstrate that our method effectively outperforms various strong baseline models.

\section{Acknowledgments}
The work described in this paper was supported by and Research Grants Council of Hong Kong (15207122, 15207920, 15207821, 15204018), National Natural Science Foundation of China (62076212) and PolyU internal grants (ZVQ0, ZVVX).

\bibliography{aaai23.bib}

\begin{thebibliography}{42}
\providecommand{\natexlab}[1]{#1}

\bibitem[{Auer et~al.(2007)Auer, Bizer, Kobilarov, Lehmann, Cyganiak, and
  Ives}]{auer2007dbpedia}
Auer, S.; Bizer, C.; Kobilarov, G.; Lehmann, J.; Cyganiak, R.; and Ives, Z.~G.
  2007.
\newblock DBpedia: {A} Nucleus for a Web of Open Data.
\newblock In \emph{6th International Semantic Web Conference, 2nd Asian
  Semantic Web Conference}, 722--735.

\bibitem[{Chen et~al.(2019)Chen, Lin, Zhang, Ding, Cen, Yang, and
  Tang}]{chen2019towards}
Chen, Q.; Lin, J.; Zhang, Y.; Ding, M.; Cen, Y.; Yang, H.; and Tang, J. 2019.
\newblock Towards Knowledge-Based Recommender Dialog System.
\newblock In \emph{Proceedings of the 2019 Conference on Empirical Methods in
  Natural Language Processing and the 9th International Joint Conference on
  Natural Language Processing, {EMNLP-IJCNLP}}, 1803--1813.

\bibitem[{Christakopoulou, Radlinski, and
  Hofmann(2016)}]{christakopoulou2016towards}
Christakopoulou, K.; Radlinski, F.; and Hofmann, K. 2016.
\newblock Towards Conversational Recommender Systems.
\newblock In \emph{Proceedings of the 22nd {ACM} International Conference on
  Knowledge Discovery and Data Mining (SIGKDD)}, 815--824.

\bibitem[{Deng et~al.(2021)Deng, Li, Sun, Ding, and Lam}]{deng2021unified}
Deng, Y.; Li, Y.; Sun, F.; Ding, B.; and Lam, W. 2021.
\newblock Unified Conversational Recommendation Policy Learning via Graph-based
  Reinforcement Learning.
\newblock In \emph{{SIGIR} '21: The 44th International {ACM} {SIGIR} Conference
  on Research and Development in Information Retrieval}, 1431--1441.

\bibitem[{Fleiss(1971)}]{fleiss1971measuring}
Fleiss, J.~L. 1971.
\newblock Measuring nominal scale agreement among many raters.
\newblock \emph{Psychological bulletin}, 76(5): 378.

\bibitem[{Gao et~al.(2021)Gao, Lei, He, de~Rijke, and Chua}]{gao2021advances}
Gao, C.; Lei, W.; He, X.; de~Rijke, M.; and Chua, T. 2021.
\newblock Advances and challenges in conversational recommender systems: {A}
  survey.
\newblock \emph{{AI} Open}, 2: 100--126.

\bibitem[{Hayati et~al.(2020)Hayati, Kang, Zhu, Shi, and
  Yu}]{hayati2020inspired}
Hayati, S.~A.; Kang, D.; Zhu, Q.; Shi, W.; and Yu, Z. 2020.
\newblock {INSPIRED:} Toward Sociable Recommendation Dialog Systems.
\newblock In \emph{Proceedings of the 2020 Conference on Empirical Methods in
  Natural Language Processing ({EMNLP})}, 8142--8152.

\bibitem[{He, Zhuo, and Law(2017)}]{he2017distributed}
He, J.; Zhuo, H.~H.; and Law, J. 2017.
\newblock Distributed-Representation Based Hybrid Recommender System with Short
  Item Descriptions.
\newblock \emph{CoRR}, abs/1703.04854.

\bibitem[{Jannach et~al.(2021)Jannach, Manzoor, Cai, and
  Chen}]{jannach2021survey}
Jannach, D.; Manzoor, A.; Cai, W.; and Chen, L. 2021.
\newblock A Survey on Conversational Recommender Systems.
\newblock \emph{{ACM} Comput. Surv.}, 54(5): 105:1--105:36.

\bibitem[{Kingma and Ba(2015)}]{kingma2014adam}
Kingma, D.~P.; and Ba, J. 2015.
\newblock Adam: {A} Method for Stochastic Optimization.
\newblock In \emph{3rd International Conference on Learning Representations
  ({ICLR})}.

\bibitem[{Lei et~al.(2020)Lei, Zhang, He, Miao, Wang, Chen, and
  Chua}]{lei2020interactive}
Lei, W.; Zhang, G.; He, X.; Miao, Y.; Wang, X.; Chen, L.; and Chua, T. 2020.
\newblock Interactive Path Reasoning on Graph for Conversational
  Recommendation.
\newblock In \emph{{KDD} '20: The 26th {ACM} {SIGKDD} Conference on Knowledge
  Discovery and Data Mining}, 2073--2083.

\bibitem[{Li et~al.(2018)Li, Kahou, Schulz, Michalski, Charlin, and
  Pal}]{li2018towards}
Li, R.; Kahou, S.~E.; Schulz, H.; Michalski, V.; Charlin, L.; and Pal, C. 2018.
\newblock Towards Deep Conversational Recommendations.
\newblock In \emph{Advances in Neural Information Processing Systems},
  9748--9758.

\bibitem[{Li et~al.(2021)Li, Lei, Wu, He, Jiang, and Chua}]{li2021seamlessly}
Li, S.; Lei, W.; Wu, Q.; He, X.; Jiang, P.; and Chua, T. 2021.
\newblock Seamlessly Unifying Attributes and Items: Conversational
  Recommendation for Cold-start Users.
\newblock \emph{{ACM} Trans. Inf. Syst.}, 39(4): 40:1--40:29.

\bibitem[{Li et~al.(2022)Li, Xie, Zhu, Ao, Zhuang, and He}]{li2022user}
Li, S.; Xie, R.; Zhu, Y.; Ao, X.; Zhuang, F.; and He, Q. 2022.
\newblock User-Centric Conversational Recommendation with Multi-Aspect User
  Modeling.
\newblock In \emph{{SIGIR} '22: The 45th International {ACM} {SIGIR} Conference
  on Research and Development in Information Retrieval}, 223--233.

\bibitem[{Liang et~al.(2021)Liang, Hu, Xu, Miao, He, Chen, Geng, Liang, and
  Jiang}]{NTRD2021EMNLP}
Liang, Z.; Hu, H.; Xu, C.; Miao, J.; He, Y.; Chen, Y.; Geng, X.; Liang, F.; and
  Jiang, D. 2021.
\newblock Learning Neural Templates for Recommender Dialogue System.
\newblock In \emph{Proceedings of the 2021 Conference on Empirical Methods in
  Natural Language Processing, {EMNLP}}, 7821--7833.

\bibitem[{Liao et~al.(2019)Liao, Takanobu, Ma, Yang, Huang, and
  Chua}]{liao2019deep}
Liao, L.; Takanobu, R.; Ma, Y.; Yang, X.; Huang, M.; and Chua, T. 2019.
\newblock Deep Conversational Recommender in Travel.
\newblock \emph{CoRR}, abs/1907.00710.

\bibitem[{Liu et~al.(2021)Liu, Wang, Niu, Wu, and Che}]{liu2021durecdial}
Liu, Z.; Wang, H.; Niu, Z.; Wu, H.; and Che, W. 2021.
\newblock DuRecDial 2.0: {A} Bilingual Parallel Corpus for Conversational
  Recommendation.
\newblock In Moens, M.; Huang, X.; Specia, L.; and Yih, S.~W., eds.,
  \emph{Proceedings of the 2021 Conference on Empirical Methods in Natural
  Language Processing ({EMNLP})}, 4335--4347.

\bibitem[{Liu et~al.(2020)Liu, Wang, Niu, Wu, Che, and Liu}]{liu2020towards}
Liu, Z.; Wang, H.; Niu, Z.; Wu, H.; Che, W.; and Liu, T. 2020.
\newblock Towards Conversational Recommendation over Multi-Type Dialogs.
\newblock In \emph{Proceedings of the 58th Annual Meeting of the Association
  for Computational Linguistics (ACL)}, 1036--1049.

\bibitem[{Lu et~al.(2021)Lu, Bao, Song, Ma, Cui, Wu, and He}]{lu2021revcore}
Lu, Y.; Bao, J.; Song, Y.; Ma, Z.; Cui, S.; Wu, Y.; and He, X. 2021.
\newblock RevCore: Review-Augmented Conversational Recommendation.
\newblock In \emph{Findings of the Association for Computational Linguistics},
  1161--1173.

\bibitem[{Ma, Takanobu, and Huang(2021)}]{ma2020cr}
Ma, W.; Takanobu, R.; and Huang, M. 2021.
\newblock CR-Walker: Tree-Structured Graph Reasoning and Dialog Acts for
  Conversational Recommendation.
\newblock In \emph{Proceedings of the 2021 Conference on Empirical Methods in
  Natural Language Processing, {EMNLP}}, 1839--1851.

\bibitem[{Manning, Raghavan, and Sch{\"{u}}tze(2008)}]{schutze2008introduction}
Manning, C.~D.; Raghavan, P.; and Sch{\"{u}}tze, H. 2008.
\newblock \emph{Introduction to information retrieval}.
\newblock Cambridge University Press.
\newblock ISBN 978-0-521-86571-5.

\bibitem[{Mikolov et~al.(2013)Mikolov, Sutskever, Chen, Corrado, and
  Dean}]{MikolovSCCD13}
Mikolov, T.; Sutskever, I.; Chen, K.; Corrado, G.~S.; and Dean, J. 2013.
\newblock Distributed Representations of Words and Phrases and their
  Compositionality.
\newblock In \emph{Advances in Neural Information Processing Systems},
  3111--3119.

\bibitem[{Papineni et~al.(2002)Papineni, Roukos, Ward, and
  Zhu}]{papineni2002bleu}
Papineni, K.; Roukos, S.; Ward, T.; and Zhu, W. 2002.
\newblock Bleu: a Method for Automatic Evaluation of Machine Translation.
\newblock In \emph{Proceedings of the 40th Annual Meeting of the Association
  for Computational Linguistics (ACL)}, 311--318.

\bibitem[{Ren et~al.(2021)Ren, Yin, Chen, Wang, Huang, and
  Zheng}]{ren2021learning}
Ren, X.; Yin, H.; Chen, T.; Wang, H.; Huang, Z.; and Zheng, K. 2021.
\newblock Learning to Ask Appropriate Questions in Conversational
  Recommendation.
\newblock In \emph{The 44th International {ACM} Conference on Research and
  Development in Information Retrieval (SIGIR)}, 808--817.

\bibitem[{Schafer et~al.(2007)Schafer, Frankowski, Herlocker, and
  Sen}]{SchaferFHS07}
Schafer, J.~B.; Frankowski, D.; Herlocker, J.~L.; and Sen, S. 2007.
\newblock Collaborative Filtering Recommender Systems.
\newblock In \emph{The Adaptive Web, Methods and Strategies of Web
  Personalization}, volume 4321, 291--324. Springer.

\bibitem[{Schlichtkrull et~al.(2018)Schlichtkrull, Kipf, Bloem, van~den Berg,
  Titov, and Welling}]{schlichtkrull2018modeling}
Schlichtkrull, M.~S.; Kipf, T.~N.; Bloem, P.; van~den Berg, R.; Titov, I.; and
  Welling, M. 2018.
\newblock Modeling Relational Data with Graph Convolutional Networks.
\newblock In \emph{The Semantic Web - 15th International Conference, {ESWC}},
  volume 10843, 593--607.

\bibitem[{Serban et~al.(2017)Serban, Sordoni, Lowe, Charlin, Pineau, Courville,
  and Bengio}]{serban2017hierarchical}
Serban, I.~V.; Sordoni, A.; Lowe, R.; Charlin, L.; Pineau, J.; Courville,
  A.~C.; and Bengio, Y. 2017.
\newblock A Hierarchical Latent Variable Encoder-Decoder Model for Generating
  Dialogues.
\newblock In \emph{Proceedings of the Thirty-First {AAAI} Conference on
  Artificial Intelligence}, 3295--3301.

\bibitem[{Speer, Chin, and Havasi(2017)}]{speer2017conceptnet}
Speer, R.; Chin, J.; and Havasi, C. 2017.
\newblock ConceptNet 5.5: An Open Multilingual Graph of General Knowledge.
\newblock In Singh, S.; and Markovitch, S., eds., \emph{Proceedings of the
  Thirty-First {AAAI} Conference on Artificial Intelligence}, 4444--4451.

\bibitem[{Sun and Zhang(2018)}]{sun2018conversational}
Sun, Y.; and Zhang, Y. 2018.
\newblock Conversational Recommender System.
\newblock In \emph{The 41st International {ACM} Conference on Research and
  Development in Information Retrieval (SIGIR)}, 235--244.

\bibitem[{Vaswani et~al.(2017)Vaswani, Shazeer, Parmar, Uszkoreit, Jones,
  Gomez, Kaiser, and Polosukhin}]{vaswani2017attention}
Vaswani, A.; Shazeer, N.; Parmar, N.; Uszkoreit, J.; Jones, L.; Gomez, A.~N.;
  Kaiser, L.; and Polosukhin, I. 2017.
\newblock Attention is All you Need.
\newblock In \emph{Advances in Neural Information Processing Systems},
  5998--6008.

\bibitem[{Wang et~al.(2021)Wang, Hu, Sha, Xu, Wong, and
  Jiang}]{wang2021finetuning}
Wang, L.; Hu, H.; Sha, L.; Xu, C.; Wong, K.; and Jiang, D. 2021.
\newblock Finetuning Large-Scale Pre-trained Language Models for Conversational
  Recommendation with Knowledge Graph.
\newblock \emph{CoRR}, abs/2110.07477.

\bibitem[{Xie et~al.(2021)Xie, Yu, Zhao, and Li}]{xie2021comparison}
Xie, Z.; Yu, T.; Zhao, C.; and Li, S. 2021.
\newblock Comparison-based Conversational Recommender System with Relative
  Bandit Feedback.
\newblock In Diaz, F.; Shah, C.; Suel, T.; Castells, P.; Jones, R.; and Sakai,
  T., eds., \emph{The 44th International {ACM} Conference on Research and
  Development in Information Retrieval (SIGIR)}, 1400--1409.

\bibitem[{Xu et~al.(2020)Xu, Moon, Liu, Liu, Shah, and Yu}]{XuMLLSY20}
Xu, H.; Moon, S.; Liu, H.; Liu, B.; Shah, P.; and Yu, P.~S. 2020.
\newblock User Memory Reasoning for Conversational Recommendation.
\newblock In \emph{Proceedings of the 28th International Conference on
  Computational Linguistics, {COLING}}, 5288--5308.

\bibitem[{Xu et~al.(2021)Xu, Yang, Xu, Gao, Guo, and Wen}]{XuYXG0W21}
Xu, K.; Yang, J.; Xu, J.; Gao, S.; Guo, J.; and Wen, J. 2021.
\newblock Adapting User Preference to Online Feedback in Multi-round
  Conversational Recommendation.
\newblock In \emph{{WSDM} '21, The Fourteenth {ACM} International Conference on
  Web Search and Data Mining}, 364--372.

\bibitem[{Zhang et~al.(2019{\natexlab{a}})Zhang, Yu, Shen, Jin, and
  Chen}]{zhang2019text}
Zhang, R.; Yu, T.; Shen, Y.; Jin, H.; and Chen, C. 2019{\natexlab{a}}.
\newblock Text-Based Interactive Recommendation via Constraint-Augmented
  Reinforcement Learning.
\newblock In \emph{Advances in Neural Information Processing Systems
  (NeurIPS)}, 15188--15198.

\bibitem[{Zhang et~al.(2019{\natexlab{b}})Zhang, Tong, Xu, and
  Maciejewski}]{zhang2019graph}
Zhang, S.; Tong, H.; Xu, J.; and Maciejewski, R. 2019{\natexlab{b}}.
\newblock Graph convolutional networks: a comprehensive review.
\newblock \emph{Computational Social Networks}, 6(1): 1--23.

\bibitem[{Zhang et~al.(2021)Zhang, Liu, Zhong, Zhang, Wang, and
  Miao}]{zhang2021kecrs}
Zhang, T.; Liu, Y.; Zhong, P.; Zhang, C.; Wang, H.; and Miao, C. 2021.
\newblock {KECRS:} Towards Knowledge-Enriched Conversational Recommendation
  System.
\newblock \emph{CoRR}, abs/2105.08261.

\bibitem[{Zhou et~al.(2021)Zhou, Wang, Zhou, Shang, Cheng, Zhao, Li, and
  Wen}]{zhou2021crslab}
Zhou, K.; Wang, X.; Zhou, Y.; Shang, C.; Cheng, Y.; Zhao, W.~X.; Li, Y.; and
  Wen, J. 2021.
\newblock CRSLab: An Open-Source Toolkit for Building Conversational
  Recommender System.
\newblock In \emph{Proceedings of the Joint Conference of the 59th Annual
  Meeting of the Association for Computational Linguistics and the 11th
  International Joint Conference on Natural Language Processing, {ACL-IJCNLP}},
  185--193.

\bibitem[{Zhou et~al.(2020{\natexlab{a}})Zhou, Zhao, Bian, Zhou, Wen, and
  Yu}]{zhou2020improving}
Zhou, K.; Zhao, W.~X.; Bian, S.; Zhou, Y.; Wen, J.; and Yu, J.
  2020{\natexlab{a}}.
\newblock Improving Conversational Recommender Systems via Knowledge Graph
  based Semantic Fusion.
\newblock In \emph{{KDD} '20: The 26th {ACM} {SIGKDD} Conference on Knowledge
  Discovery and Data Mining}, 1006--1014.

\bibitem[{Zhou et~al.(2020{\natexlab{b}})Zhou, Zhao, Wang, Wang, Zhang, Wang,
  and Wen}]{zhou2020leveraging}
Zhou, K.; Zhao, W.~X.; Wang, H.; Wang, S.; Zhang, F.; Wang, Z.; and Wen, J.
  2020{\natexlab{b}}.
\newblock Leveraging Historical Interaction Data for Improving Conversational
  Recommender System.
\newblock In \emph{{CIKM} '20: The 29th {ACM} International Conference on
  Information and Knowledge Management}, 2349--2352.

\bibitem[{Zhou et~al.(2020{\natexlab{c}})Zhou, Zhou, Zhao, Wang, and
  Wen}]{zhou2020towards}
Zhou, K.; Zhou, Y.; Zhao, W.~X.; Wang, X.; and Wen, J. 2020{\natexlab{c}}.
\newblock Towards Topic-Guided Conversational Recommender System.
\newblock In \emph{Proceedings of the 28th International Conference on
  Computational Linguistics ({COLING})}, 4128--4139.

\bibitem[{Zhou et~al.(2022)Zhou, Zhou, Zhao, Wang, Jiang, and Hu}]{zhou2022c2}
Zhou, Y.; Zhou, K.; Zhao, W.~X.; Wang, C.; Jiang, P.; and Hu, H. 2022.
\newblock C{\({^2}\)}-CRS: Coarse-to-Fine Contrastive Learning for
  Conversational Recommender System.
\newblock In \emph{{WSDM} '22: The Fifteenth {ACM} International Conference on
  Web Search and Data Mining}, 1488--1496.

\end{thebibliography}

\end{document}